\documentclass[rotating]{aa}
\usepackage{amsmath}
\usepackage{txfonts}
\usepackage[final]{epsfig} 
\usepackage{epsf}
\usepackage[final]{graphics}
\usepackage{natbib}
\usepackage[american,english]{babel}

\begin{document}
\title{Small-scale structure of the galactic cirrus emission 
%
  \thanks{Based on observations with ISO, an ESA project with instruments
    funded by ESA Member States (especially the PI countries: France,
    Germany, the Netherlands and the United Kingdom) and
    with the participation of ISAS and NASA.}
  }

\author{ Cs.~Kiss\inst{1,2}
  \and  P.~\'Abrah\'am\inst{1,2} 
  \and  U.~Klaas\inst{1}
  \and  D.~Lemke\inst{1}  
  \and  Ph.~H\'eraudeau\inst{1}
  \and  C.~del~Burgo\inst{1}
  \and  U.~Herbstmeier\inst{1} }
\institute{ Max-Planck-Institut f\"ur Astronomie, K\"onigstuhl~17,
     D-69117~Heidelberg, Germany
  \and  Konkoly Observatory of the Hungarian Academy of Sciences, 
    P.O. Box 67, H-1525~Budapest, Hungary
  }
\offprints{Cs.~Kiss, pkisscs@konkoly.hu}
\date{ Received  / Accepted ...}
%
\abstract{We examined the Fourier power spectrum 
characteristics of cirrus structures in 13 sky fields with faint to bright 
cirrus emission observed with ISOPHOT in the 90--200$\mu$m wavelength range in order 
to study variations of the spectral index $\alpha$.
We found that $\alpha$ varies from field to field 
with --5.3\,$\le$\,$\alpha$\,$\le$\,--2.1.
It depends on the absolute surface brightness and on the
hydrogen column density. We also found 
different spectral indices for the same sky region 
at different wavelengths. Longer wavelength measurements show  
steeper power spectra. This can be explained by the presence of
dust at various temperatures, in particular of a cold extended component. 
For the faintest areas of the far-infrared sky 
we derive a wavelength independent spectral index of
$\alpha$\,=\,--2.3$\pm$0.6  for the cirrus power spectrum.
The application of the correct spectral index is a precondition
for the proper disentanglement of the cirrus foreground component
of the Cosmic Far-Infrared Background and its fluctuations. 

}
\maketitle
\section{Introduction\label{introduction}}
\begin{figure*}
\hbox{\hskip 1.1cm \epsfxsize=4.8cm 
		\epsffile{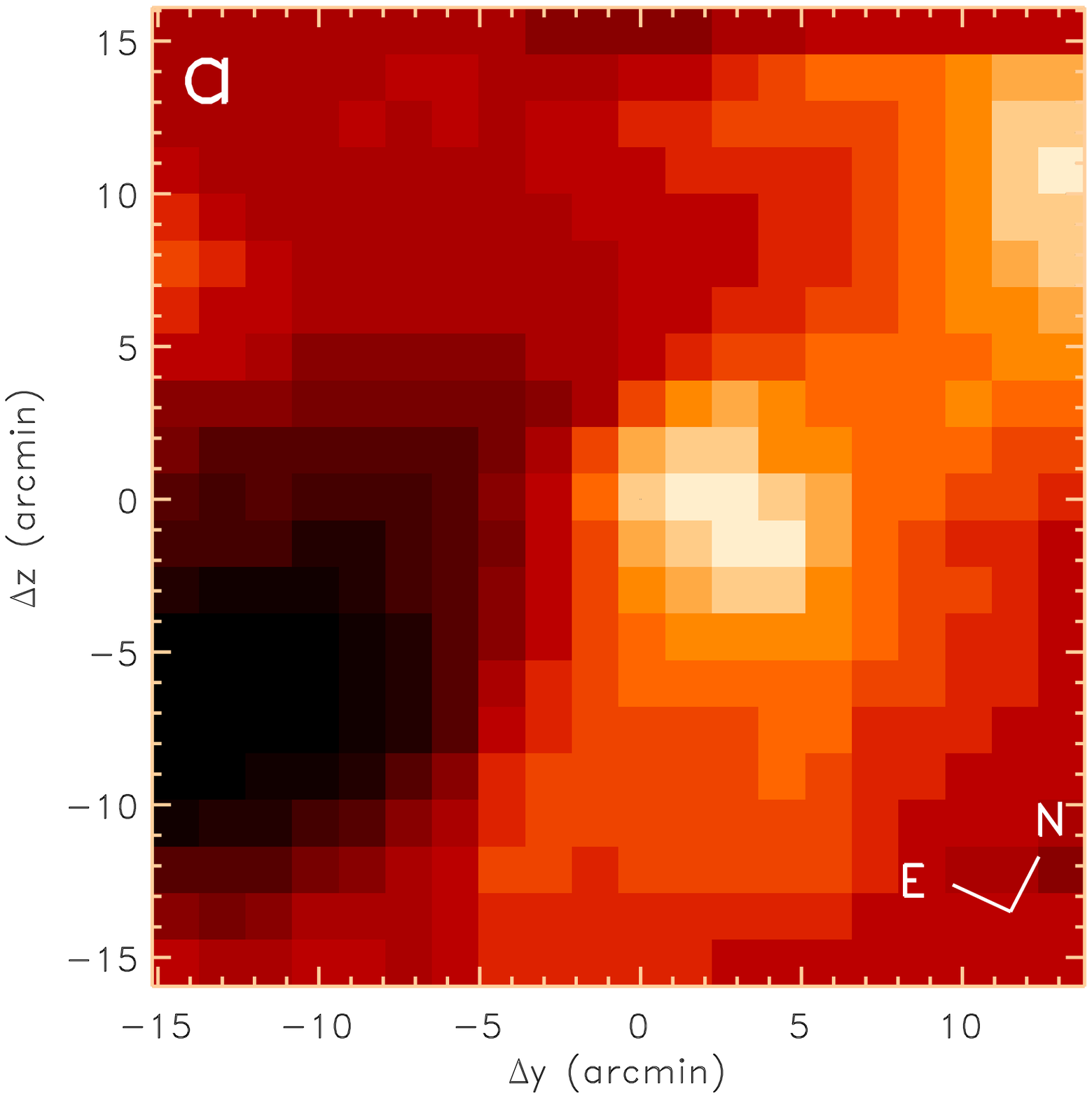}
	\hskip 1.2cm \epsfxsize=4.8cm 
		\epsffile{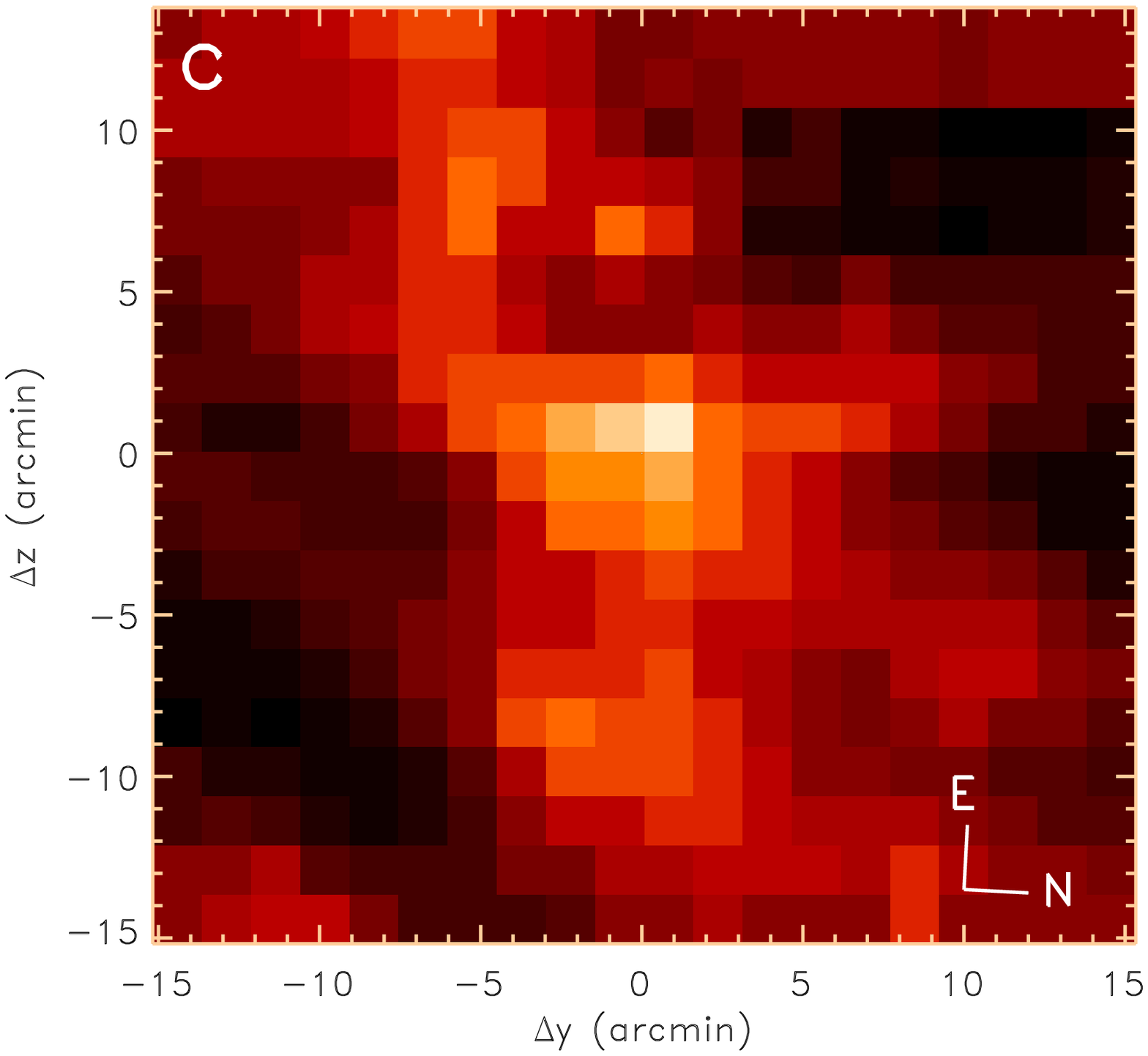}}
\vskip 0.8cm	
\hbox{\hskip 1.1cm \epsfxsize=4.8cm 
		\epsffile{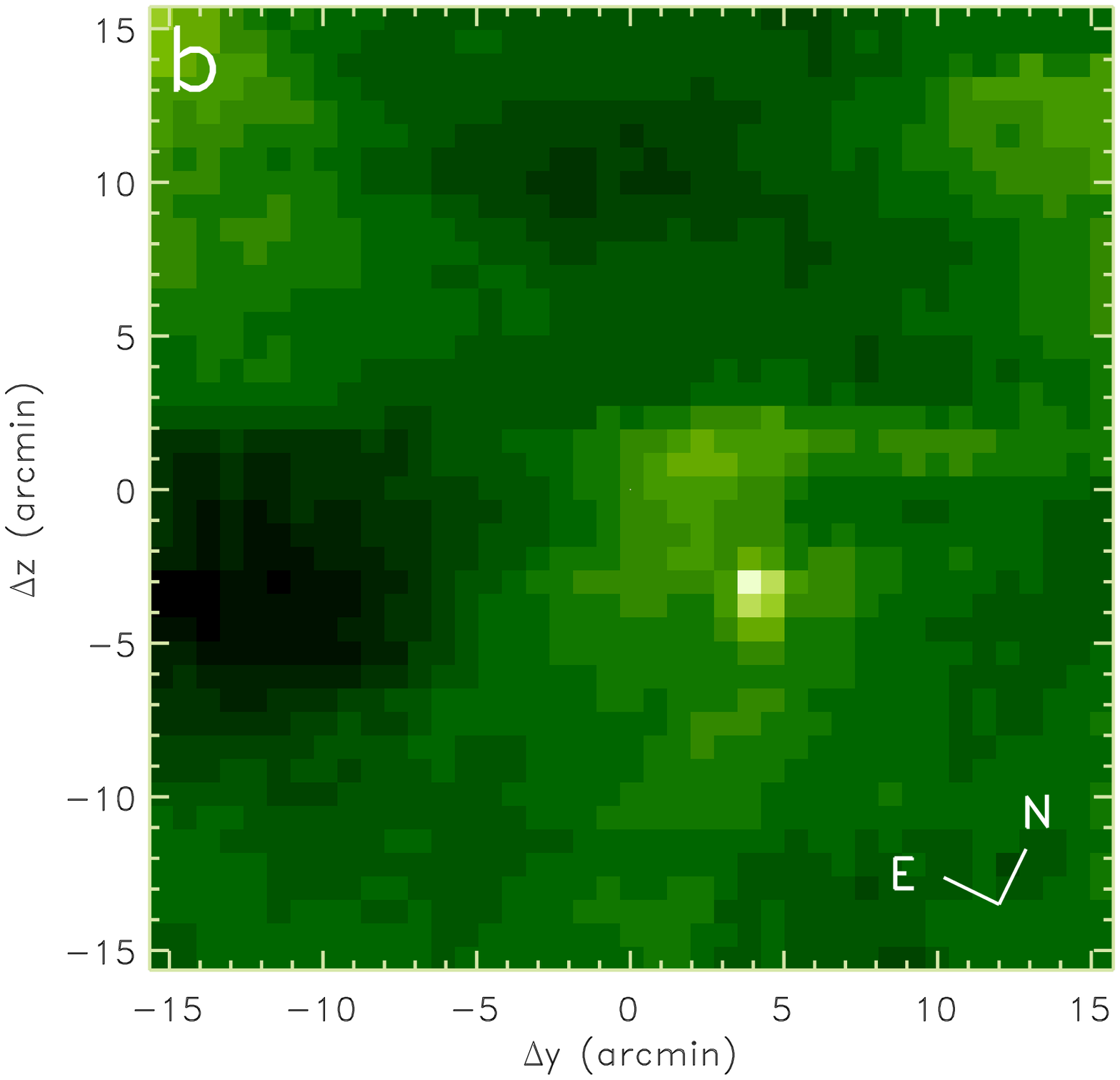}
	\hskip 1.2cm \epsfxsize=4.8cm 
		\epsffile{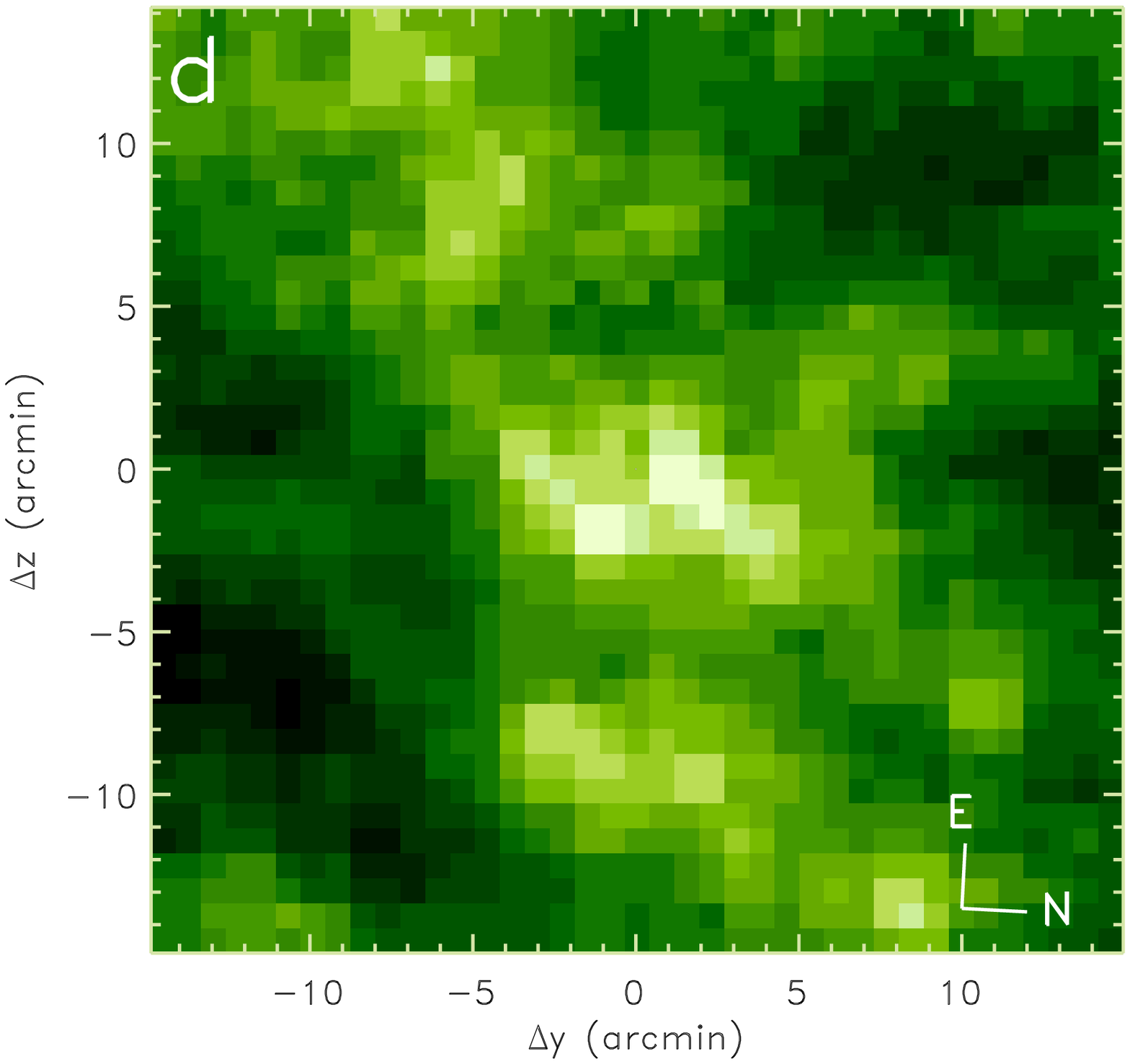}}	
\vskip -5.8cm 
\hfill
\parbox[b]{48mm}{
\caption[]{100 and 200\,$\mu$m range images of two sample fields
in the satellite coordinate system \citep{Laureijs2001}.
The central sky positions of the images are listed in Table~1.
The orientation of the celestial coordinate system is indicated.
({\bf a}) Cham--II, 200\,$\mu$m
({\bf b}) Cham--II, 100\,$\mu$m
({\bf c}) NPC\,1, 200\,$\mu$m
({\bf d}) NPC\,1, 90\,$\mu$m. 
While the FIR emission in NPC\,1 arises from the same cirrus
structure at both wavelengths, there seems to be an additional
component of cold condensations in Cham\,II, as 
discussed in Sect.~\ref{sect:disc_alphavar}.  }
\label{fig:images}}
\vskip 0.5cm
\end{figure*}
In the far-infrared the most significant component
superimposed on the Extragalactic Background  is the cirrus emission of 
the Galaxy originating from irregularly shaped interstellar clouds \citep{Low_84}.  
The correct determination of the Cosmic Far-InfraRed Background 
(CFIRB) requires a precise removal of this foreground component
{\citep{Guiderdoni}}. 
The CFIRB is characterized by two parameters: the absolute level 
and the amplitude of its fluctuations. 
In recent years observations made by ISOPHOT 
\citep{Lemke96}, the photometer
on board the ISO satellite \citep{Kessler96}
made the determination of the CFIRB fluctuations possible.

CFIRB fluctuations are expected to show the following properties
{\citep[see e.g.][]{Hauser+Dwek}}:
(1) the spatial distribution corresponds to a flat power spectrum,
(2) they are isotropic and 
(3) they have a positive constant value. 
In contrast, the fluctuations due to galactic cirrus 
emission scale up with absolute brightness and can be described 
by a multifractal structure, yielding a steep Fourier power spectrum.
It is usually described by one main parameter,
the spectral index $\alpha$:
\begin{equation}
\rm P = P_0 \times (f/f_{0})^\alpha
\label{eq:powerlaw}
\end{equation}
where P is the Fourier power,
f is the absolute value of the two-dimensional spatial frequency and
P$_0$ is the Fourier power at the spatial frequency f$_0$
\citep{Gautier}.
Since cirrus fluctuations are different from CFIRB fluctuations in 
at least two 
features (spatial frequency-- and brightness--dependence), it is 
possible to separate them from each other in two independent ways: 

(1) The decomposition through the brightness dependence at a fixed 
spatial frequency was performed by 
\citet[][hereafter Paper I]{Kiss2001},
and resulted in a new determination of the CFIRB fluctuation 
amplitudes at 90 and 170\,$\mu$m. 

(2) The decomposition using the Fourier power spectrum was introduced
by \citet{Guiderdoni}. 
For faint fields, where CFIRB fluctuations
are expected to be dominant in the highest spatial frequency range,
the knowledge of the cirrus spectral index is mandatory for
decomposition. 
\citet{Gautier} determined a spectral index of the 
galactic cirrus emission $\alpha$\,$\approx$\,--3
from the analysis of 100\,$\mu$m IRAS scans.
For regions of faint total FIR brightness emission \citet{Lagache2000}
and \citet{Matsuhara} derived CFIRB fluctuation
amplitudes applying this spectral index of $\alpha$\,$\approx$\,--3 
for the cirrus decomposition.

\citet{Herbstmeier} analysed a few fields
observed by ISOPHOT in the 90--180\,$\mu$m wavelength range. 
They found a variation of spectral indices in the range 
of --0.5\,$\ge$\,$\alpha$\,$\ge$\,--3.6, depending on wavelength,
brightness and sky position. However, this sample
was too small (four fields, two of them measured only at one
wavelength) for general conclusions, but their results clearly
indicated that the $\alpha$\,=\,--3 cirrus spectral index may not
be universal. 

Despite the strong impact of the cirrus power spectrum 
on the correct determination of the characteristics of the
CFIRB, cirrus power spectrum properties were not yet analysed
in detail for $\lambda$\,$>$\,100$\mu$m and at a
resolution achievable by ISOPHOT. 
In this paper we analyse 13 sky regions measured by 
ISOPHOT in the 90--200\,$\mu$m range and 
investigate the characteristics of their Fourier power spectra.

\section{Observations and data analysis \label{sect:observations}}

\subsection{Selection of ISOPHOT maps \label{sect:selection}}

We selected 20 FIR maps from the ISO archive, covering 13
separate sky regions. All selected maps were obtained in the PHT22 staring 
raster observing mode \citep{Laureijs2001}. 
We excluded maps containing strong point sources or obvious spatial structures. 
This selection is a subsample of the fields used in Paper I.
The selected maps had to show a clear sign of cirrus emission 
in the power spectrum. This requires a minimum average surface brightness of 
$\langle$B$\rangle$\,$\approx$\,3\,MJysr$^{-1}$.
Such fields are slightly brighter than the faintest fields in the
far-infrared sky ($\langle$B$\rangle$\,$\approx$\,2\,MJysr$^{-1}$)
in the 90--200\,$\mu$m wavelength range 
after the subtraction of the Zodiacal Light.

Our sample covers the surface brightness range
$\sim$\,3--60\,MJysr$^{-1}$, ranging from   
faint to bright cirrus, and faint molecular cloud
emission. The basic properties of the maps are 
summarized in Table~1. Images of two fields are shown in Fig.\ref{fig:images},
more images are presented in \citet{Herbstmeier}.

\begin{table*}[ht!]
\small
\hskip 2cm
\rotatebox[]{90}{ 
\begin{tabular}{lccccccccccccc} 
\hline \small
(1) & (2) & (3) & (4) & (5) & (6) & (7) & (8) & (9) & (10) &
(11) & (12) & (13) & (14) \\ \hline
field & $\lambda$ & size   & \multicolumn{2}{c}{field center} & 
        $\rm \langle B_\lambda \rangle$ & $\alpha$ & P$_0$ & 
        $\delta\alpha$ & ${\delta}P_0$  & W($^{12}$CO) &
	N(HI) & associated objects & references \\
  &  ($\mu$m) & (arcmin) & l & b & (MJysr$^{-1}$) &  & (Jy$^2$sr$^{-1}$) & 
     &   &  (Kkms$^{-1}$) & (cm$^{-2}$) &  &  \\ \hline   
NGP    & 180  & 46.0$\times$46.0  &  88.9 &  73.0 & 3.1$\pm$0.1
       & --2.53$\pm$0.22 
       & 1.8$\times$10$^3$ & 1.59 & 0.95 & $-$ & 9.0$\times$10$^{19}$ & 
       North Galactic Pole & \citet{Juvela} \\
       & & & & & & & & & & & & & \cite{Herbstmeier} \\
Draco  & 90   & 7.7$\times$7.7    &  89.8 &  38.6 & 5.7$\pm$0.3
       & --2.45$\pm$0.17 
       & 5.3$\times$10$^3$ & 1.47 & 0.76 & $-$ & 2.1$\times$10$^{20}$ & 
       Draco nebula & \citet{Herbstmeier} \\
Draco  & 170  & 27.6$\times$18.4  &  89.8 &  38.6 & 4.7$\pm$0.5
       & --4.09$\pm$0.22 & 2.9$\times$10$^3$ &  
        2.71  & 1.39 & -- & 2.1$\times$10$^{20}$ &
        Draco nebula &  \citet{Herbstmeier} \\
Cep    & 90   & 14.7$\times$20.7  & 108.0 &  13.6 & 23.5$\pm$3.1
       & --3.42$\pm$0.15 
       & 3.3$\times$10$^4$ & 1.96 & 1.01 & 6.9 & 2.5$\times$10$^{21}$ 
       & Cepheus Flare & \citet{Herbstmeier} \\
Cep    & 170  & 21.5$\times$24.5  & 108.0 &  13.6 & 68.2$\pm$12.7
       & --4.19$\pm$0.12       
       & 3.0$\times$10$^5$ & 1.95 & 1.56 & 6.9 & 2.5$\times$10$^{21}$ & 
       Cepheus Flare &  \citet{Herbstmeier}\\
NPC1   & 90   & 29.9$\times$29.1  & 121.6 &  24.2 & 7.6$\pm$0.8
       & --3.10$\pm$0.04 
       & 5.0$\times$10$^3$ & 2.16 & 0.64 & 6.6 & 6.9$\times$10$^{20}$ & 
       Polaris Flare & \citet{Gautier} \\
NPC1   & 200  & 30.7$\times$29.1  & 121.6 &  24.2 & 26.1$\pm$5.3
       & --3.46$\pm$0.09 
       & 1.1$\times$10$^5$ & 2.30 & 0.95 & 6.6 & 6.9$\times$10$^{20}$ & 
       Polaris Flare & \citet{Gautier} \\
NPC2   & 90   & 29.9$\times$29.1  & 122.0 &  24.6 & 7.4$\pm$0.7  
       & --2.83$\pm$0.09 
       & 5.7$\times$10$^3$ & 1.74 & 0.89 & 5.5 & 6.7$\times$10$^{20}$ & 
       Polaris Flare & \citet{Gautier} \\
NPC2   & 200  & 30.7$\times$29.1  & 122.0 &  24.6 & 22.6$\pm$3.8
       & --3.39$\pm$0.08 
       & 9.7$\times$10$^4$ & 2.19 & 0.63 & 5.5 & 6.7$\times$10$^{20}$ & 
       Polaris Flare & \citet{Gautier} \\
North1 & 90   & 29.9$\times$29.9  & 100.0 &  14.8 & 13.3$\pm$1.4
       & --3.05$\pm$0.09 
       & 9.7$\times$10$^3$ & 1.83 & 0.91 & 4.5 & 1.6$\times$10$^{21}$ 
       & \object{LDN\,1122} & \citet{Yonekura} \\
North1 & 200  & 30.7$\times$30.7  & 100.0 &  14.8 & 34.4$\pm$5.6 
       & --4.70$\pm$0.17 
       & 4.5$\times$10$^4$  & 2.43 & 0.52 & 4.5 & 1.6$\times$10$^{21}$ 
       & \object{LDN\,1122} & \citet{Yonekura} \\
North2 & 90   & 29.9$\times$29.9  & 108.0 &  15.2 & 12.4$\pm$1.8
       & --3.65$\pm$0.13 
       & 9.2$\times$10$^3$ & 1.42 & 0.90 & 6.6 & 2.1$\times$10$^{21}$ 
       & \object{LDN\,1147, 1148} & \citet{Lee} \\
       & & & & & & & & & & & & & \citet{Yonekura} \\
North2 & 200  & 30.7$\times$30.7  & 108.0 &  15.2 & 40.7$\pm$12.5
       & --3.81$\pm$0.09 
       & 3.2$\times$10$^5$ & 1.94 & 0.61 & 6.6 & 2.1$\times$10$^{21}$ 
       & \object{LDN\,1147, 1148} & \citet{Lee} \\
       & & & & & & & & & & & & & \citet{Yonekura} \\
M01    & 180  & 27.6$\times$27.6  & 100.0 &  30.6 & 5.0$\pm$0.2
       & --2.56$\pm$0.41 
       & 2.3$\times$10$^2$ & 0.89 & 0.75 & $-$ & 4.2$\times$10$^{20}$ 
       & faint field in Dra & \citet{Herbstmeier}\\
       & & & & & & & & & & & & & \cite{Abraham} \\
M03    & 180  & 27.6$\times$27.6  & 117.6 &  46.1 & 3.7$\pm$0.1
       & --2.12$\pm$0.23 
       & 5.8$\times$10$^2$ & 0.82 & 0.71 & $-$ & 1.5$\times$10$^{20}$ 
       & faint field in UMi & \cite{Abraham} \\
TMC2-1 & 200  & 30.7$\times$30.7  & 173.9 & --15.7& 56.5$\pm$8.9
       & --5.26$\pm$0.15 
       & 3.7$\times$10$^5$ & 1.74 & 0.86 & 18.9 & 1.7$\times$10$^{21}$ 
       & \object{LDN\,1529, 1531} & \citet{Lee} \\
       & & & & & & & & & & & & Taurus\,2 mol. cl. & \citet{Scalo}\\
TMC2-2 & 200  & 30.7$\times$30.7  & 174.3 & --15.9& 63.7$\pm$8.5 
       & --3.49$\pm$0.14 
       & 3.8$\times$10$^5$ & 1.66 & 1.04 & 16.3 & 1.8$\times$10$^{21}$ 
       & \object{LDN\,1529, 1531} & \citet{Lee}\\
       & & & & & & & & & & & & Taurus\,2 mol. cl. &  \citet{Scalo}\\
Cha1S  & 100  & 19.9$\times$19.9  & 297.3 & --16.2& 29.3$\pm$3.2 
       & --3.80$\pm$0.08 
       & 1.4$\times$10$^5$ & 2.47 & 0.79 & 10.3 & 9.2$\times$10$^{20}$ & 
       Chamaeleon mol. cl.  & 
      \citet{Toth} \\
ChamII & 100  & 31.4$\times$31.4  & 303.5 & --14.2& 23.9$\pm$3.3
       & --2.95$\pm$0.08 
       & 1.7$\times$10$^5$ & 0.91 & 0.89 & 16.1 & 1.2$\times$10$^{21}$ 
       & DCld\,303.3--14.3, & \citet{Toth} \\ 
ChamII & 200  & 30.7$\times$30.7  & 303.5 & --14.2& 38.4$\pm$14.6
       & --5.17$\pm$0.18 
       & 2.4$\times$10$^5$ & 2.48 & 0.53 & 16.1 & 1.2$\times$10$^{21}$ & 
       303.5--14.4 & \cite{Toth} \\ \hline
\end{tabular}}
\normalsize
\vskip -23.7cm 
\hskip 13.5cm 
\rotatebox{90}{
\begin{minipage}{23cm} \small {\bf Table 1.}
Basic observational and derived parameters of fields mapped by ISOPHOT. 
The columns are: (1) name of the field
as specified by the observer; (2) central wavelength of filter; 
(3) map size; (4)--(5) center of the field in galactic coordinates; 
(6) median surface brightness from ISOPHOT photometry with Zodiacal 
Light subtracted; quoted standard deviations are a measure of the
dynamic range of the map. 
(7) spectral index of the power spectrum; 
(8) fluctuation power at the reference spatial frequency 1/d$_0$
(d$_0$\,=\,4\arcmin); 
(9)--(10) parameters describing the relative strength of the instrument
noise power spectrum to the signal power spectrum 
(see Eq.~2 for the definitions); 
(11) average neutral hydrogen column density \citep{Dickey+Lockman};
(12) $^{12}$CO integrated intensity \citep{Dame};
(13) associated objects or dark clouds located 
(at least partially) inside the field.
(14) references with the description of the field. 
\end{minipage}}
\clearpage
\label{table-main}
\end{table*}

\subsection{Data reduction \label{sect:datareduction}}
 
We used the final maps produced
for the confusion noise analysis in Paper~I. For a detailed description 
of the data reduction steps we refer to this paper.
The data reduction comprised the following main steps: 
  \begin{itemize}
    \item Basic data analysis with 
    PIA\footnote{PIA is a joint develpoment 
    by the ESA Astrophysics Division and the ISOPHOT consortium led by the
    Max-Planck-Institut f\"ur Astronomie (MPIA), Heidelberg}~V8.2
    \citep{Gabriel} from ERD to AAP level
    \item Flat-fielding using first quartile normalization
    \item Subtraction of the Zodiacal emission 
  \end{itemize}

\subsection{Fourier power spectrum \label{sect:data_fourier}}
  
Fourier transformation of the maps was performed by routines 
written in IDL\footnote{Interactive Data Language, Version 5.2 and 5.3, 
Research Systems Inc.} based on the standard IDL FFT routine with 
periodogram normalisation \citep{Press}.  
{For all maps the highest spatial sampling frequency corresponds to the
pixel size of the C100 and C200 array cameras. The Nyquist limit which
allows correct sampling (without aliasing) is half of that frequency
\citep[][Chapter~12.1]{Press}, i.e. the double pixel size of the array. 
Therefore, the upper limit of the spatial frequency considered for the
Fourier power spectrum was set to the Nyquist-limit, 
which is $\theta_{min}$\,=\,92\arcsec~
for images with $\lambda$\,$\le$\,100\,$\mu$m (C100 detector)
and $\theta_{min}$\,=\,184\arcsec~for images with 
$\lambda$\,$>$\,100\,$\mu$m (C200 detector). 
Since the pixel sizes of the ISOPHOT C100 and C200 cameras 
were adjusted to the diffraction limit,
this limit should be kept, even if the raster sampling were finer than the
pixel size.  }

Instead of averaging P(f) fluctuation power values in annuli
for each f\,=\,$|${\bf f}$|$ spatial frequency, we used all individual 
data points (f, P(f) pairs) to derive the final power spectrum. 
{The spectral index $\alpha$ is derived by robust line fitting
to all data points in the log(f)--log(P) space.}

In most cases the whole power spectrum could be satisfactorily 
fitted by one spectral index. 
For some faint fields, however, the high frequency end
of the spectrum showed a different (flatter) spectral index,
due to a higher {structure noise} level.
{In such cases we only used the low and mid frequency part to derive
$\alpha$. The reason for this flattening will be discussed 
below in Sect.~4. }

\section{Features affecting the power spectrum}

There are several contributions to the power spectrum
that can affect its final shape.
We analyse the effects of the instrument
noise, a point-source footprint and an extended source
profile on the final power spectrum.   
The correct separation of the power spectrum components 
requires well-distinguishable spectral indices, 
otherwise they cannot be disentangled.

\subsection{Power spectrum of the instrument noise}


\begin{figure}
\epsfig{file=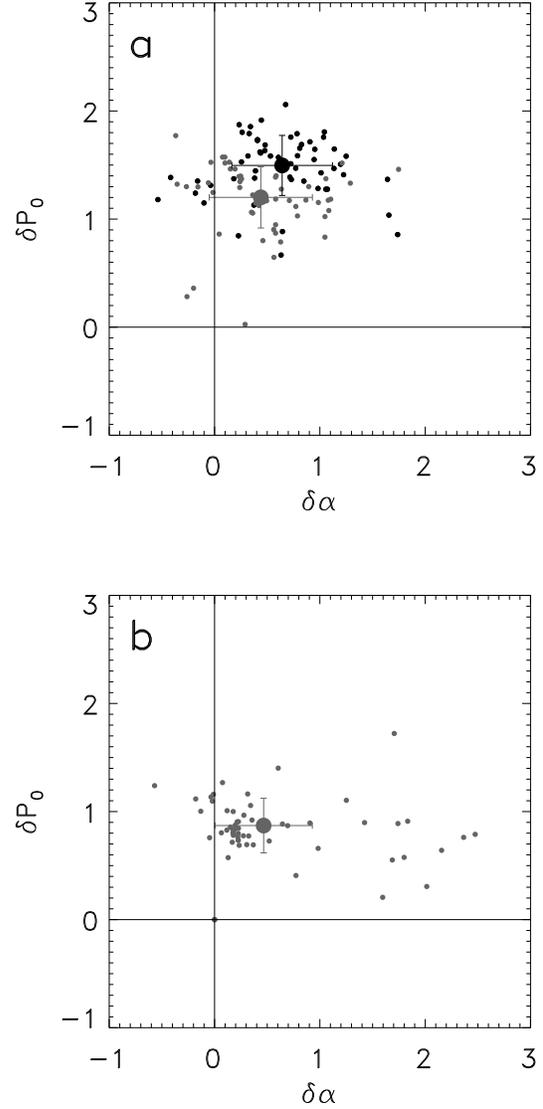, width=8.5cm}
\caption{Relative strength of the instrument noise- to the signal
power spectrum. 
(a) Subsample of 56 maps from Paper~I., observed with the C200 detector
in oversampling mode.
Black dots: flat-field noise; gray dots: PIA-noise. 
(b) Subsample of 60 maps from Paper~I., observed with the
C100 detector.
In this case no representative values of the 
flat-field noise could be dervied due to the lack 
of a sufficient number of oversampled maps.
The average values and 
the standard deviations both in $\delta\alpha$ and $\delta{P_0}$ are
marked via large dots and error bars, respectively. 
}
\label{fig:instpowers}
\end{figure}
Instrument noise can be a significant contribution 
to the fluctuation power at low surface brightness and  
high spatial frequencies. As discussed in detail in Paper~I, 
the most representative estimates of the instrument noise 
are PIA-noise (which 
reflects the statistical uncertainties in signal derivation)
and flat-field noise (which reflects the variation of the signal from
individual detector pixels at the same sky position).
Flat-field noise can only be 
determined for oversampled maps. 

We assume that both the signal and instrument noise 
power spectra - derived from the surface brightness maps and the
corresponding noise maps, respectively - are described by a power law behaviour 
as in Eq.~\ref{eq:powerlaw}.
%
%
Here the $\rm f_0$ values correspond to the Nyquist-limits, 
i.e$\theta_{min}$\,=\,92\arcsec~ and
184\arcsec~ for the C100 and C200 detectors, respectively.

The signal power spectrum is not affected by the instrument noise, if  
the power spectrum of the instrument noise is reasonably weaker
than that of the signal in the frequency range analysed. 
This is ensured, if two criteria are fulfilled simultaneously: 
(1) the fluctuation power
of the instrument noise P$_i$ is lower by at least a 
factor of 2 than that of the signal 
P$s$ at the resolution limit, $\rm\theta_{min}$ and
(2) the power spectrum of the signal is steeper than that of the
instrument noise. 
These conditions are parametrized by the following two
quantities ($i$ and $s$ denote instrument noise and signal, respectively): 
\begin{equation}
\rm\delta\alpha = \alpha_i - \alpha_s  > 0
~~~~~~
\delta{P_0} = log_{10} \bigg( {{P_0^s}\over{P_0^i}} \bigg) > 0.3
\end{equation} 
%


{For the instrument noise analysis we used a larger sample of
maps than for the cirrus power spectrum study. In Table~1
we provide the respective values of $\delta\alpha$ and $\rm\delta P_0$
for our special fields, which can be compared with the ranges of 
$\delta\alpha$ and $\rm\delta P_0$ for the larger sample of maps.}

In the case of the C200 detector we calculated 
$\delta\alpha$ and $\delta{P_0}$ for 
a subsample of 56 maps from Paper I,  
observed with full oversampling (Fig.~\ref{fig:instpowers}a). 

Most of the flat-field and PIA-noise data points
are located in the quadrant where both $\delta\alpha$ and $\delta{P_0}$ 
are positive. Although there are a few points in the negative
$\delta\alpha$ regime, they are close to zero, 
representing nearly parallel instrument noise and signal power spectra.
Since in all cases $\delta{P_0}$\,$>$\,0.3, 
the instrument noise power spectrum does not exceed the signal 
power spectrum for frequencies below $\rm\theta_{min}^{-1}$.  
The data points with $\alpha\,\la\,0$ belong to 
the faintest regions of the sky with a nearly flat 
power spectrum, showing no signs of cirrus 
($\rm\alpha_s$ and $\alpha_i \approx$\,0). 
The distributions of the PIA- and flat-field noise data points 
are rather similar, and the average values are within
the standard deviations in $\delta\alpha$ and $\delta{P_0}$
(big black and gray dots with error bars in Fig.~\ref{fig:instpowers}a).
This demonstrates that the method of noise determination has no
impact on the final result. 

In the case of the C100 detector it was not possible to perform a similar 
investigation due to the lack of suitable oversampled maps. 
Therefore, we analysed 60 observations from Paper~I
by calculating the PIA-noise values only (Fig.~\ref{fig:instpowers}b).
Also the C100 detector measurements fulfil the
requirements of a suitable signal-to-noise ratio. 

This noise analysis justifies, that for both C200 and C100 maps
of Table~1 instrument noise does not affect the final shape
of the power spectrum, in particular since for all of them
$\delta\alpha$\,$>$\,0.8 and $\delta{P_0}$\,$>$\,0.5. 

\rm 

\rm
\subsection{Power spectra of point- and extended sources
 \label{sect:pp_extended}}
%
\begin{figure}
\epsfig{file=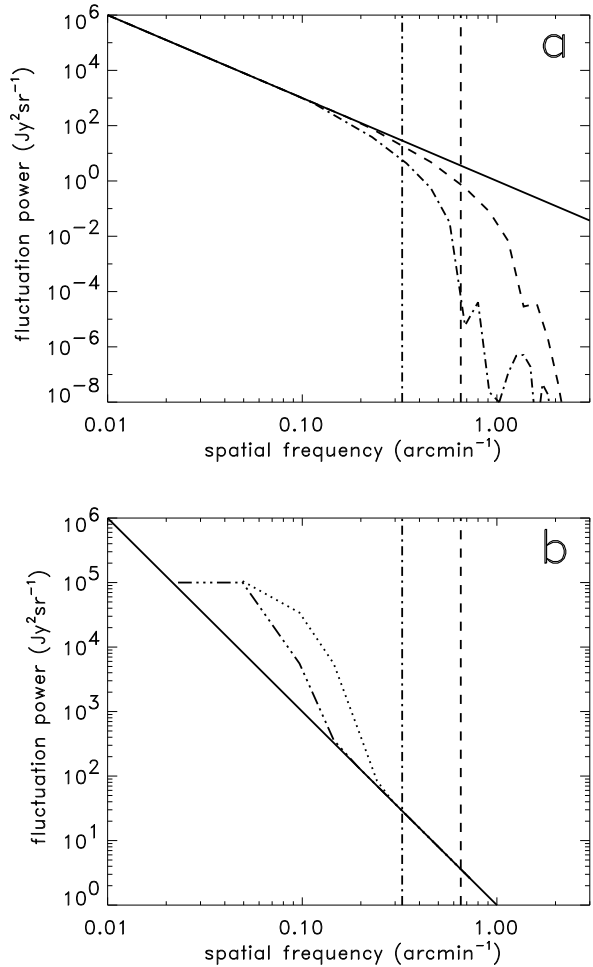, width=8.5cm, angle=0}
\caption[]{Schematic representation of the contribution of
different emission components to the final power spectrum.
({\bf a}) Convolution of a theoretical cirrus spectrum of arbitrary 
stength ($\alpha$\,=\,--3, P\,=\,10$^3$\,Jy$^2$sr$^{-1}$ at 
f\,=\,0.1\,arcmin$^{-1}$)
with ISOPHOT's theoretical point-source footprint at 
90\,$\mu$m (dashed line) and at 200\,$\mu$m
(dash-dotted line). The pure,
non-convolved cirrus spectrum is also shown (solid line).
({\bf b}) Resulting power spectrum of a cirrus-like intensity 
distribution ($\alpha$\,=\,--3, solid line) and an extended
source of Gaussian shape with a FWHM of 1/6 (dotted line) and
1/3 (dash--triple-dotted line) of the size of the original map.
The Gaussians are normalized to P\,=\,10$^5$\,Jy$^2$sr$^{-1}$
at f\,=\,0\,arcmin$^{-1}$. 
The Nyquist-limits for the C100 and C200 filter maps are
marked both in (a) and (b) by vertical dashed and dash-dotted lines, respectively. 
}
\vskip 0.4cm
\label{fig:modelpp}  
\end{figure}

\begin{figure*}
\centerline{\epsfig{file=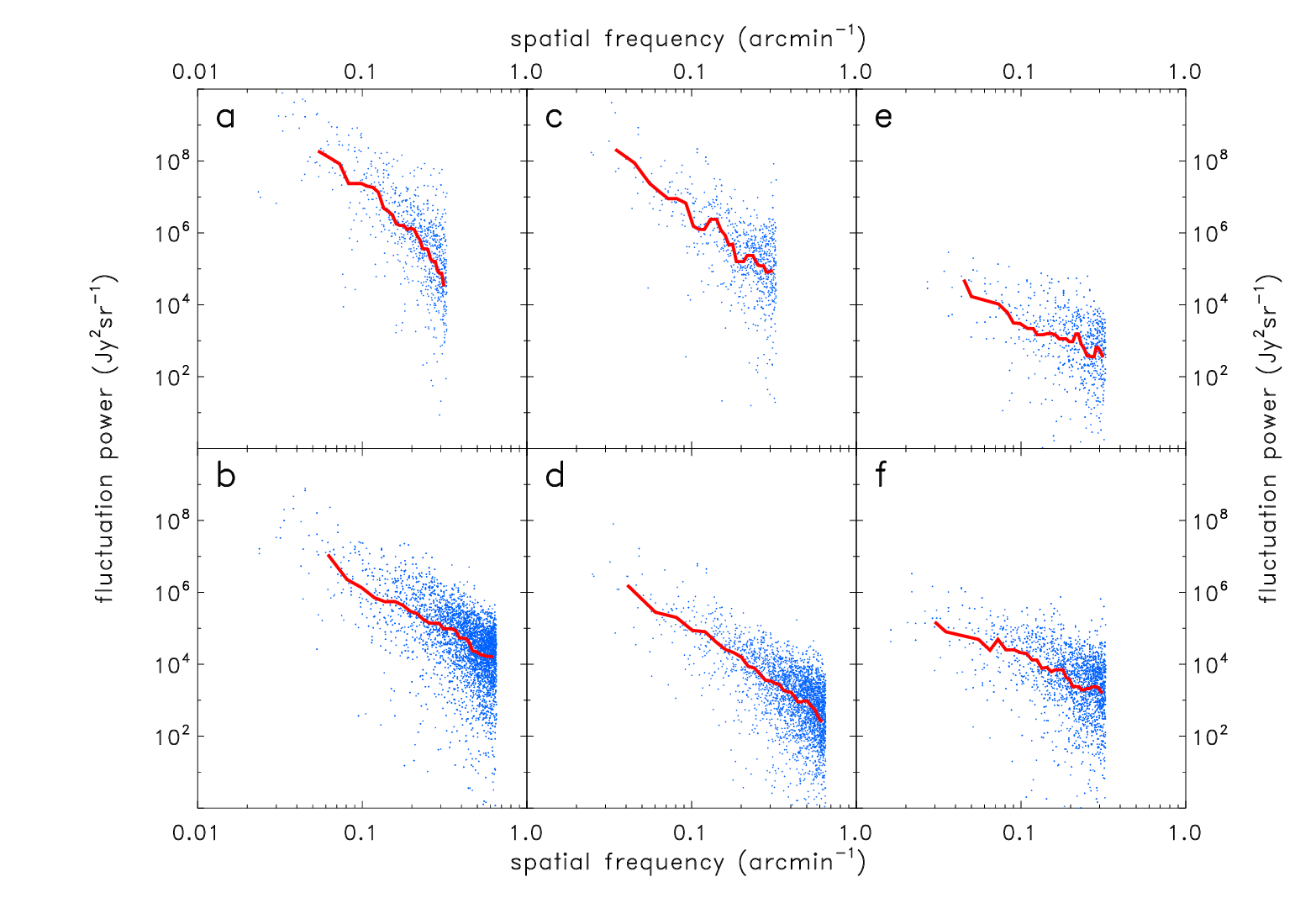, width=17cm, angle=0}}
\caption[]{Power spectra of six sample fields, two of them observed by both the 
C100 and the C200 camera. 
The images corresponding to the first four power spectra 
are presented in Fig.~\ref{fig:images}. Images of M01 and NGP can be
found in \citet{Herbstmeier}. 
({\bf a}) Cham--II, 200\,$\mu$m
({\bf b}) Cham--II, 100\,$\mu$m
({\bf c}) NPC\,1, 200\,$\mu$m
({\bf d}) NPC\,1, 90\,$\mu$m
({\bf e}) M01, 180\,$\mu$m
({\bf f}) NGP, 180\,$\mu$m. 
{Each dot represents a (f,P(f)) pair, the solid line is smoothed
average to outline the trend. The high frequency end is determined 
by the Nyquist-limit, the low frequency end by the size of the map.}
} 
\label{fig:power}  
\end{figure*}
In Fig.~\ref{fig:modelpp}a we present a model cirrus spectrum 
with an expected cirrus spectral index of $\rm \alpha\,=\,-3$ 
convolved with the 90\,$\mu$m and 200\,$\mu$m theoretical 
footprints 
\citep[footprint\,=\,convolution of the telescope point spread function with the pixel aperture,][]{Laureijs2001}. 
This figure shows, that the footprint 
has only a little effect below the actual Nyquist-limit, and this effect
can be neglected when deriving the spectral index. A strong point source
may dominate the whole spectrum, but maps containing these features
were not included in our sample. 

As will be shown in Sect.~5.1, extended sources with excess emission may be
superimposed on the cirrus-like intensity distribution. We tested the
effect of these sources on the power spectrum assuming a Gaussian shape,
and a FWHM of 1/3 and 1/6 of the size of the original map, respectively, 
as presented in Fig.~\ref{fig:modelpp}b. 
The shape of the resulting power spectrum strongly depends on the 
relative strength of the cirrus and extended source components.  
The typical effect of the extended source is the appearance of a 
'hump' at medium spatial frequencies, if the strengths of the
two components are comparable. Similar (however weaker) features 
can be identified in some of our real power spectra, especially at
200\,$\mu$m (see Fig.~\ref{fig:power}). 
The high frequency part
of an extended source power spectrum is rather steep and therefore
-- if dominant -- can lead to a power spectrum steeper than
the one of pure cirrus. 
A set of extended sources has a power spectrum similar to 
that of a single one, but the higher the number of sources,
the shallower the final power spectrum. In certain configurations 
these spectra may be described by an average spectral index of 
$\alpha$\,$\approx$\,--3, 
leading to a spectrum, which is in practice undistinguishable from 
the one of cirrus.

\section{Results \label{sect:results}}

Table~1 and Fig.~\ref{fig:power} 
present the results of the power spectrum analysis.

The fields in our analysis were characterized by the following
physical parameters: median FIR surface brightness 
from ISOPHOT photometry with Zodiacal Light subtracted
(this work), neutral hydrogen column density \citep{Dickey+Lockman} 
and $^{12}$CO integrated intensity \citep{Dame}.
The power spectra follow quite well the expected power-law behaviour
up to the spatial frequency corresponding to the resolution limit,
and therefore can be satisfactorily fitted by one single
spectral index $\alpha$. {The only exceptions are the fields
NGP (Fig.~\ref{fig:power}f), M01 (Fig.~\ref{fig:power}e) and M03 at 180\,$\mu$m. 
They show a flattening of the spectrum for the highest spatial 
frequencies which was eventually excluded in the derivation of the
$\alpha$ values given in Table~1.
}

{We suggest that the relatively strong 
structure noise at high spatial frequencies 
arises from the fluctuations
due to the extragalactic background. As discussed in Paper~I, 
the extragalactic background is the major contributor to the structure 
noise at high spatial frequencies for the faintest regions of the 
far-infrared sky measured with the C200 detector. 
These three faint fields have an average surface brightness
comparable to that of the cosmological fields. 
The extragalactic background is expected to have a fluctuation 
power of $\rm\sim{8\times}10^2$\,$Jy^2sr^{-1}$ at 170\,$\mu$m with our 
FFT normalization. Fluctuation powers at the reference spatial 
frequency listed in Table~1 (column \#8) are fitted values
using the cirrus-like power spectrum, therefore they can be lower than the expected
CFIRB fluctuation level.}

We calculated the median values of the spectral indices for the C100 and 
C200 fields ($\rm\overline{\alpha}_{C1}$
and $\rm\overline{\alpha}_{C2}$, respectively),
in order to compare them with previous results. 
This resulted in 
$\rm\overline{\alpha}_{C1}~=~-3.15{\pm}0.48$ and
$\rm\overline{\alpha}_{C2}~=~-3.87{\pm}1.06$.
The median C100 spectral index is not much different from the one determined
by \citet{Gautier} from the analysis of IRAS
100$\mu$m scans, however it shows a relatively large scatter.
{\citet{Abergel} performed a pixelwise determination of the
spectral index for a 12\fdg5$\times$12\fdg5 ISSA map close to the 
South Ecliptic Pole. At 100\,$\mu$m they obtained an average spectral index
of $\rm\overline{\alpha}_{100}\,=\,-3.34$ (using the relation 
$\rm\alpha\,=\,-2(\beta-1)$ between their $\beta$-value and 
our spectral index). The distribution of the spectral index over the field
had a Gaussian shape with $\sigma_\alpha$\,=\,0.36. This means, that within 
one field the variation is comparable to the variation we found for 
our sample fields.}
The C200 spectral indices are significantly steeper 
with an even higher scatter than the C100 spectral indices. 

\begin{figure}
\epsfig{file=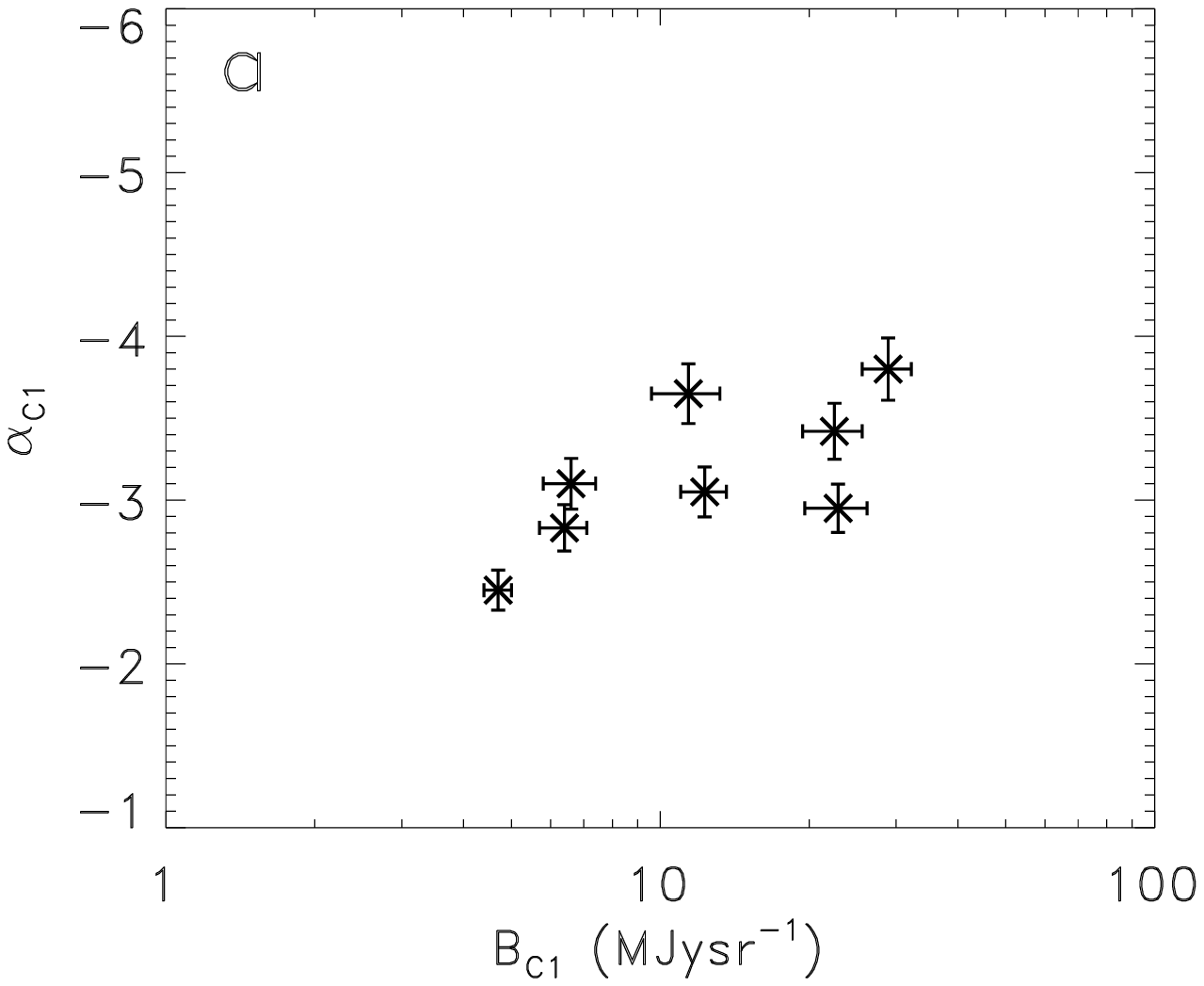, width=8.5cm}
\epsfig{file=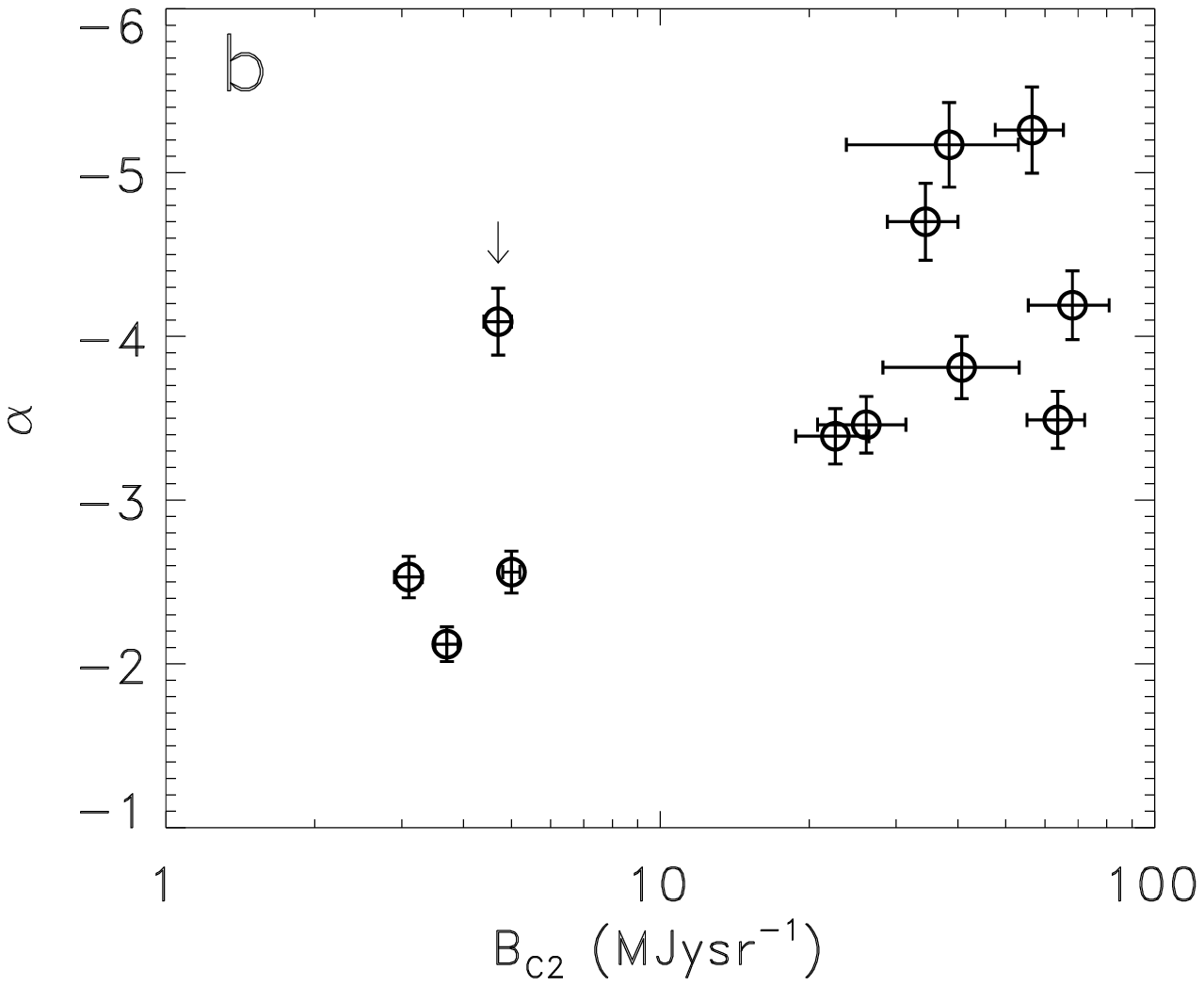, width=8.5cm}
\epsfig{file=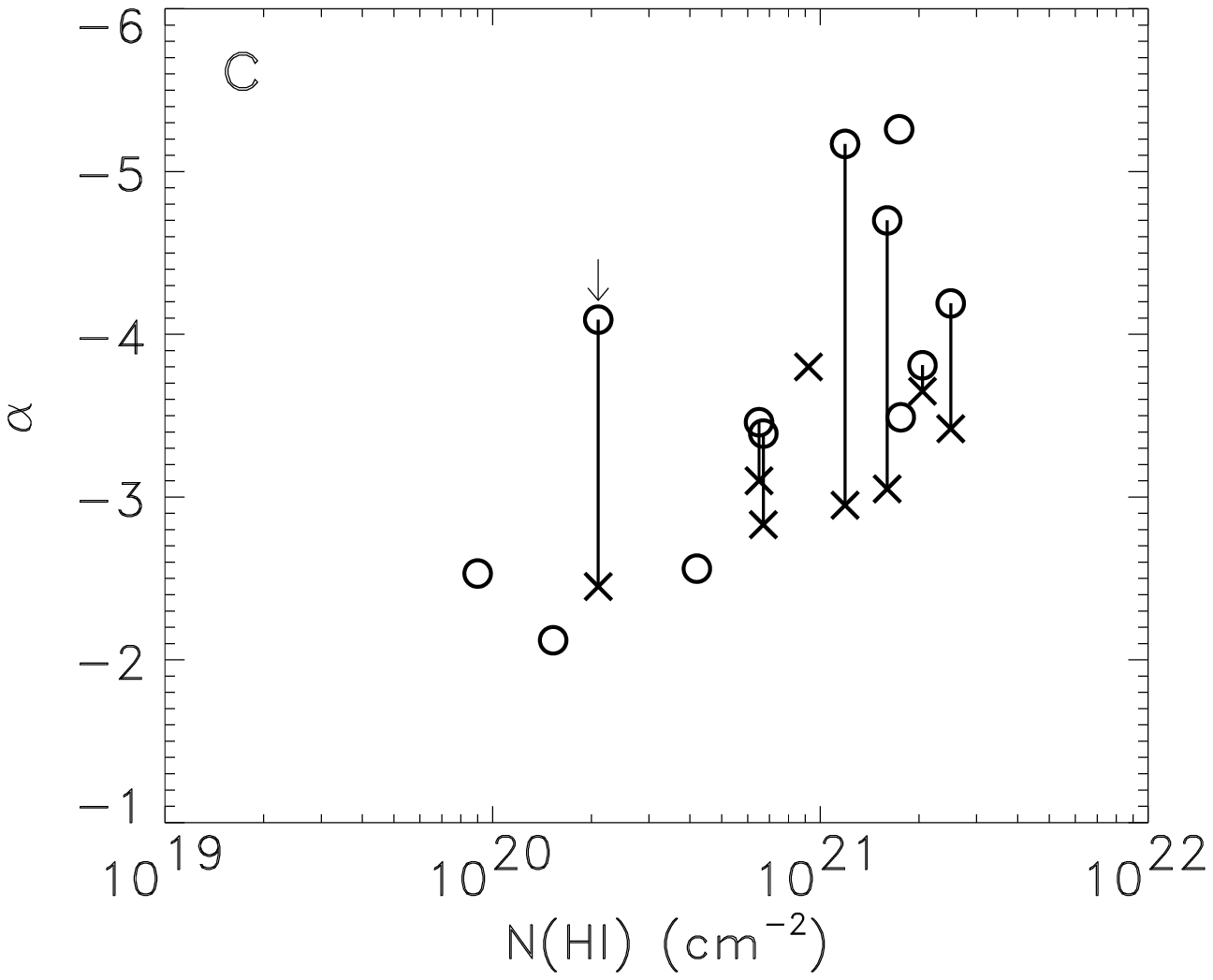, width=8.5cm}
\caption[]{Relationship of the spectral index with 
({\bf a}) the 100\,$\mu$m range (90--100\,$\mu$m)
median surface brightness B$\rm_{C1}$,
({\bf b}) the 200\,$\mu$m range (170--200\,$\mu$m)
median surface brightness B$\rm_{C2}$,
({\bf c}) the average neutral hydrogen column density.
Crosses and open circles mark fields observed at 
$\lambda$\,$\le$\,100\,$\mu$m and 
170\,$\le$\,$\lambda$\,$\le$\,200\,$\mu$m, respectively.
The data points of the Draco 170\,$\mu$m field are marked by arrows.
Error bars in (a) and (b) represent the standard deviations in B$\rm_{C1}$ and
B$\rm_{C2}$ and they mark the errors in fitting the 
spectral index in $\alpha$.}   	      
\label{fig:B-alpha}
\end{figure}

Spectral indices show a clear dependence both on the surface 
brightness and on the neutral hydrogen column density. 
Brighter fields and fields with higher HI column 
density usually have a steeper power spectrum (Fig.~\ref{fig:B-alpha}).
{We tested the degree of correlation by calculating the 
linear Pearson correlation coefficient which resulted in 0.69, 
0.80 and 0.67 for the 
$\rm log_{10} \langle B_{C1} \rangle$--$\rm\alpha_{C1}$,
$\rm log_{10} \langle B_{C2} \rangle$--$\rm\alpha_{C2}$ and
$\rm log_{10} N(HI)$--$\alpha$ relations, respectively, without the
data points of the Draco region (marked in Fig.~\ref{fig:B-alpha} by 
an arrow).  

If we exclude in Fig.~5b the three data points with 
high B$_{C2}$ and $\alpha$\,$<$\,--4.5, then the  
$\rm log_{10} \langle B_{C2} \rangle$--$\rm\alpha_{C2}$
correlation is much tighter with a linear Pearson correlation 
coefficient of 0.94. This exclusion is justified by the fact, that
all three fields have a high HI column density and two of them show a 
large $\alpha_{C2}$/$\alpha_{C1}$ ratio (Fig.~5c), indicative of a relatively 
large molecular hydrogen content, as discussed below in Sect.~5.2.
} 

{It is possible to determine an $\alpha_{C2}$ 
value for the faintest fields of the far-infrared sky, 
by a robust linear extrapolation of the 
$\rm log_{10} \langle B_{C2} \rangle$--$\rm\alpha_{C2}$ relation.
In order to account for the contribution by the CFIRB the
B$_{C2}$ values were reduced beforehand by 0.8\,MJysr$^{-1}$ (Paper~I). 
For all points except the Draco region this resulted in
$\alpha_{C2}$\,=\,--2.3$\pm$0.6 for the faintest cirrus fields
($\langle B \rangle$\,$\approx$\,2\,MJysr$^{-1}$).
Excluding in addition the three points with $\alpha$\,$<$\,--4.5,
$\alpha_{C2}$\,=\,--2.1$\pm$0.4. }
The impact of this power law in separating the CFIRB from the galactic
cirrus is discussed in Sect.~\ref{sect:disc_cfirb}.
   
\begin{figure}[h!]
\epsfig{file=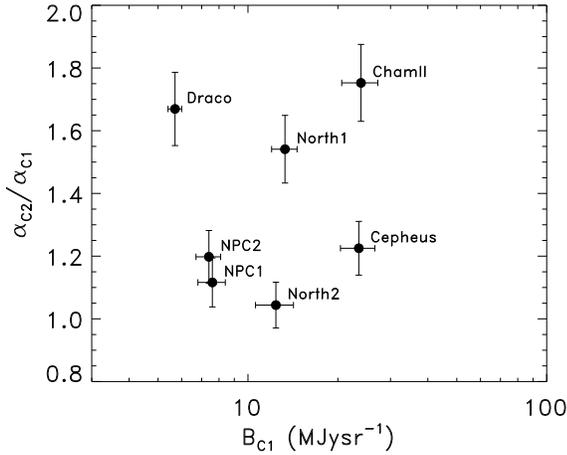, width=8.5cm}
\caption[]{Ratio of the spectral indices of sky regions 
measured in both a long wavelength (170-200\,$\mu$m, $\rm\alpha_{C2}$) 
and at short wavelength (90--100\,$\mu$m, $\rm\alpha_{C1}$) filter
versus the average surface brightness in the C100 filter 
(90 or 100$\mu$m, B$\rm_{C1}$). 
The regions are labelled by the names as given by the observer and 
as listed in Table~1.}
\label{fig:a2-a1}
\end{figure}

In Fig.~\ref{fig:a2-a1} we plot the ratio of the spectral indices 
$\rm\alpha_{C2}$/$\alpha_{C1}$ determined both for a C200 and a C100
map of the same field.   
Assuming that the FIR radiation at 100 and 200\,$\mu$m is 
emitted by the same structures the ratio $\rm \alpha_{C2} / \alpha_{C1}$
should be $\approx$\,1.   
It is obvious, that $\rm \alpha_{C2} / \alpha_{C1}$ varies in the 
range 1.0--1.8. Since the determination of the spectral index is 
quite accurate (see Table~1), this deviation
cannot be explained by measurement uncertainties. 
We propose an explanation in Sect.~\ref{sect:disc_alphavar}
 
\section{Discussion \label{sect:discussion}}
\subsection{Wavelength dependence of the spectral index
  \label{sect:disc_alphavar}}   
\begin{figure}[h!]
\epsfig{file=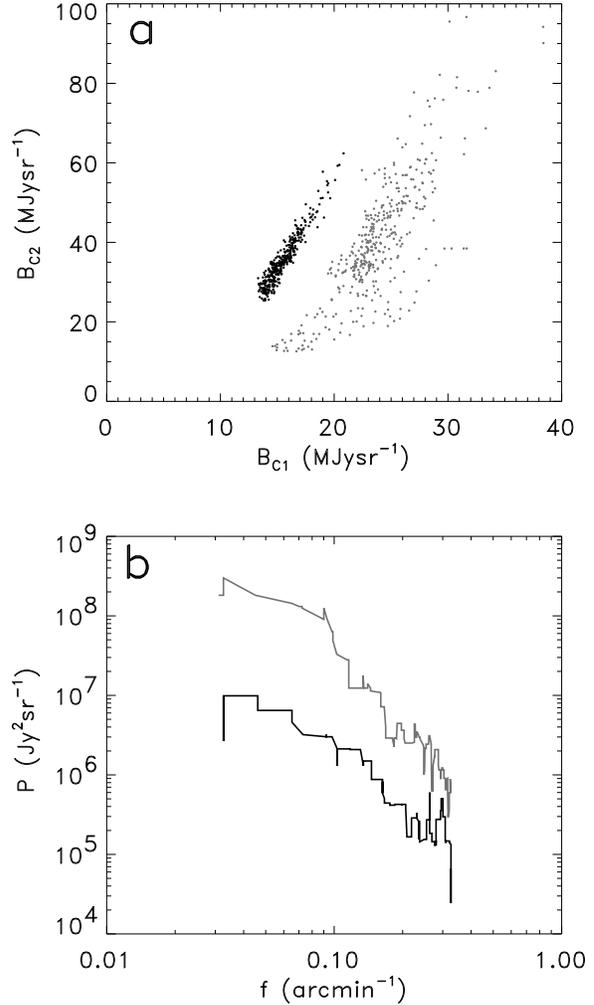, width=8.5cm}
\caption[]{Examples to demonstrate the effect of a non-uniform 
color temperature within the same field. 
({\bf a}) C100 vs. C200 surface brightness scatter plots. The 
slope is closely related to the color temperature (see text for
details). Black and gray dots represent the data points of the North\,1 and 
Cham\,II fields, respectively.
({\bf b}) Power spectra of the $\Delta$B images (cirrus component subtracted)
of Cham\,II (gray) and North\,1 (black).}
\label{fig:tempcomp}
\end{figure}

Galactic cirrus emission is expected to have a spectral energy
distribution of a modified black body, with $\nu^2$ emissivity
law, and a dust colour temperature of 17.5$\pm$1.5\,K 
\citep{Lagache98}. 
If the FIR emission
of a certain cloud can be well described by a single dust temperature, 
then the emission measured at two separate wavelengths should be 
strongly correlated, showing the same spatial structure and the same power spectrum. 
On the other hand, the presence of another 
component with different colour temperature
can lead to an excess intensity at one wavelength. This yields also 
a different spatial structure of the emission, and therefore
different power spectra at two wavelengths.
{Indications for such a dichotomy in cloud hierarchy were found
by \citet{Abergel} from their analysis of the 60 and 100\,$\mu$m IRAS
maps of the South Ecliptic Pole field. Uncorrelated features come from 
the coldest regions of the analysed fields.}
   
We tested the uniqueness of the colour temperatures for fields,
where measurements at two wavelengths were available. 
The colour temperature is represented by the flux ratio 
of the two wavelengths.
Two examples are presented in Fig.~\ref{fig:tempcomp}. 
The 90--100\,$\mu$m surface brightness maps were
smoothed to the resolution of the 170--200\,$\mu$m measurements in order 
to produce B$_{C1}$ -- B$_{C2}$ pixel-to-pixel scatter plots,
where B$_{C1}$ and B$_{C2}$ are the surface brightnesses of the
90--100\,$\mu$m and 170--200\,$\mu$m map pixels, respectively. 
In most cases the scatter plot showed a well correlated
B$_{C1}$ -- B$_{C2}$ emission and could be fitted by one straight line, 
corresponding to one single dust temperature. 
Using this relation we derived an excess 
surface brightness $\Delta$B map, by subtracting the 
scaled up B$_{C1}$ emission from B$_{C2}$. 
\begin{equation}
 \Delta B\,=\,B_{C2} - [A_1 \times B_{C1} + A_0]
\label{eq:DB}
\end{equation}
where A$_1$ and A$_0$ are the slope and the interception
of the linear fit to the scatter plot, respectively. 

However, as presented in Fig.~\ref{fig:tempcomp}a, in some cases there
was more than one characteristic colour temperature present in a field.
The most conspicuous 
example is the Cham\,II field (gray dots in Fig.~\ref{fig:tempcomp}a).
In this case, after removing the cirrus contribution 
according to the procedure of Eq.~\ref{eq:DB}, the excess surface
brightness image $\Delta$B still showed the steep index of the 
original map.
%
The steep $\Delta$B power 
spectrum can be associated to a bright extended source, as suggested 
by the examples in 
Sect~\ref{sect:pp_extended} and, indeed, the region 
contains dark cloud cores, namely 
\object{DCld\,303.3-14} and \object{DCld\,303.5-14.4} 
(see Table~1). 
The expected flattening at low spatial frequencies due to the
extended source power spectrum (see Sect.~\ref{sect:pp_extended}) can
also be identified.  The characteristic size of the
extended source derived from the power spectrum agrees well
with the typical cold core extent of $\sim$10\arcmin~ found by 
\citet{Toth} in Chamaeleon, with temperatures of $\le$\,15\,K.  
A similar effect can be 
observed for the North\,1 field, however, the differences are  
smaller than in the case of the Cham\,II region. 
The feature in the power spectrum can be identified with the 
molecular ($^{13}$CO) core of the dark cloud \object{LDN\,1122}
\citep{Yonekura}. 
In the case of the NPC\,1 field we found that the power spectrum 
of the excess surface brightness image is significantly
shallower than that of the 200\,$\mu$m surface brightness image,
because this lacks any dark cloud core. 

No colour temperature could be derived for the Draco 
region via the scatter plot procedure, due to its small spatial extent
and narrow surface brightness range. However,  
the presence of a relatively bright extended source is obvious 
in the 170$\mu$m image \citep[see][]{Herbstmeier}. 
This produces very likely the very steep power spectrum 
observed at this wavelength, despite the low average surface 
brightness (see Fig.~\ref{fig:B-alpha}), 
as already mentioned by \citet{Herbstmeier}. 

These findings suggest 
that the wavelength dependence of the 
spectral indices is caused by the combined effects of non-uniform
dust temperatures and the presence of extended sources in the
fields we selected for our study. 
The separation of the cirrus and extended source components in the
power spectra can be more easily performed by including spatial scales larger than
the characteristic extent of our maps. 
At 170\,$\mu$m such an analysis would be possible by
utilizing the ISOPHOT Serendipity Survey slews with a typical 
length of a few degrees \citep{Bogun}.

\subsection{The effect of atomic-to-molecular phase transition}

The correlation of the spectral indices 
with surface brightness and with neutral hydrogen 
column density is obvious for faint fields, while a scatter
in $\alpha$ is observed for bright and high HI column density fields,
as presented in Fig.~\ref{fig:B-alpha}.
These fields also show strong molecular ($^{12}$CO) emission,
as presented in Table~1.
This large scatter can be observed for 
N(HI)\,$\ge$\,10$^{21}$\,cm$^{-2}$, which is 
approximately the column density of the atomic-to-molecular
phase transition in the galactic cirrus \citep{deVries,Meyerdierks}.
In these clouds the ratio of molecular hydrogen column density
to atomic hydrogen column density may vary in the range of 
N(H$_2$)/N(HI)\,$\approx$\,0.2--5.0 \citep{MBM}.
Therefore clouds showing the same N(HI) may have different total 
hydrogen masses. This diverse molecular content -- 
and therefore diverse final structure -- can lead to the observed
scatter in the spectral indices. 

\subsection{Implication of cirrus power spectra on the CFIRB
fluctuations \label{sect:disc_cfirb}}


The spectral index for faint fields observed in the 170--200\,$\mu$m 
wavelength range is greater than the usually assumed $\alpha$\,=\,--3, 
which was originally derived by \citet{Gautier} from 100\,$\mu$m 
IRAS scans of medium-brightness cirrus fields. 
Due to the lack of a better determination this was 
widely applied to the faintest fields of the 
far-infrared sky and for longer wavelengths 
\citep[e.g.][]{Guiderdoni,Lagache2000}. 
We showed that brighter extended sources -- which are
often present in cirrus fields -- may significantly increase the
steepness of the power spectrum. Therefore it is plausible,
that the $\alpha$\,=\,--3 power law found for medium 
surface brightness fields results from the superposition 
of the power spectrum of a less steeper fractal structure with 
extended sources.
Our analysis provides $\alpha$\,=\,--2.3$\pm$0.6 
for the spectral index of the faintest cirrus fields. 
Although there is a discrepancy between $\alpha$-values determined at different 
wavelengths for brighter fields, the spectral indices should be the
same for faint fields, since these lack additional extended sources
and multiple dust colour temperatures. 
{We finally conclude } that a uniform $\alpha$\,=\,--2.3$\pm$0.6 cirrus spectral index
-- independent of wavelength -- has to be applied 
for the faintest areas ($\langle$B$\rangle$\,$\approx$\,2\,MJysr$^{-1}$) 
of the far-infrared sky.
The application of this $\alpha$-value would decrease the CFIRB fluctuation 
amplitudes obtained by previous works in the 170--200\,$\mu$m range
by 5--20\% (Kiss et al., 2002, in prep.).     
For somewhat brighter regions it is mandatory to determine the proper
spectral index from the properties of the local structures. 

\section{Conclusions \label{sect:conclusions}}
We examined the Fourier power spectrum 
characteristics of 13 sky fields with faint to bright cirrus emission
in the 90--200$\mu$m wavelength range and derived the
spectral index of the power spectrum, $\alpha$.
We found that $\alpha$ varies from field to field.
It has a clear dependence on the absolute surface brightness and on the
hydrogen column density of the field. 
For longer wavelengths ($\lambda$\,$>$\,100\,$\mu$m)
the scatter in $\alpha$ is larger, especially for higher 
hydrogen column densities. 
This is because for N(HI)\,$\ge$\,10$^{21}$\,cm$^{-2}$ the atomic-to-molecular
phase transition leads to
various $\alpha$ values especially in the 
170--200\,$\mu$m range, where the diverse molecular content is 
easier to observe.
We also found a significant difference in spectral indices for
the same sky region at different wavelengths. At longer wavelengths 
($\lambda\,>$\,100\,$\mu$m) the power spectra are steeper. 
This can be explained
by the coexistence of dust components with various temperatures 
within the same field and cold extended emission features
which significantly affect the power spectrum.
A spectral index of the cirrus power spectrum with a value 
of $\alpha$\,=\,--2.3$\pm$0.6 -- independent of wavelength 
-- could be derived for the faintest areas of the far-infrared sky.
The precise determination of the spectral index for each 
field will enable to correctly disentangle the galactic 
foreground components of the CFIRB. 

\rm

\begin{acknowledgements}
\sloppy\sloppy
The development and operation of ISOPHOT were supported by MPIA and
funds from Deut\-sches Zentrum f\"ur Luft- und Raumfahrt 
(DLR). The ISOPHOT Data Center at MPIA is supported
by Deut\-sches Zentrum f\"ur Luft- und Raumfahrt e.V. (DLR) with
funds of Bundesministerium f\"ur Bildung und Forschung, 
grant~no.~50~QI~0201. 
{This research was partly supported by the
ESA PRODEX programme (No.~14594/00/NL/SFe) and by the Hungarian
Research Fund (OTKA, No. T037508). P.~\'A. acknowlegdes the
support of the Bolyai Fellowship. We are indebted to the referee
Dr.\,R.\,J.~Laureijs for useful comments, which improved the paper.}

\end{acknowledgements}



\end{document}